\documentclass[a4,conference,compsocconf]{IEEEtran}
\usepackage{cite}\usepackage{graphicx}\usepackage{url}

\begin{document}

\title{Open Data:\\Reverse Engineering and Maintenance Perspective}
\author{
  \IEEEauthorblockN{Holger M. Kienle}
  \IEEEauthorblockA{hkienle@acm.org\\
    \url{http://holgerkienle.wikispaces.com/}
  }}
\maketitle


\begin{abstract}
  Open data is an emerging paradigm to share large and diverse
  datasets---primarily from governmental agencies, but also from other
  organizations---with the goal to enable the exploitation of the data
  for societal, academic, and commercial gains. There are now already
  many datasets available with diverse characteristics in terms of
  size, encoding and structure. These datasets are often created and
  maintained in an ad-hoc manner. Thus, open data poses many
  challenges and there is a need for effective tools and techniques to
  manage and maintain it.

  In this paper we argue that software maintenance and reverse
  engineering have an opportunity to contribute to open data and to
  shape its future development. From the perspective of reverse
  engineering research, open data is a new artifact that serves as
  input for reverse engineering techniques and processes. Specific
  challenges of open data are document scraping, image processing, and
  structure/schema recognition. From the perspective of maintenance
  research, maintenance has to accommodate changes of open data
  sources by third-party providers, traceability of data
  transformation pipelines, and quality assurance of data and
  transformations. We believe that the increasing importance of open
  data and the research challenges that it brings with it may possibly
  lead to the emergence of new research streams for reverse
  engineering as well as for maintenance.
\end{abstract}

\begin{figure*}[bt]
  \centering
  \includegraphics[width=1.5\columnwidth]{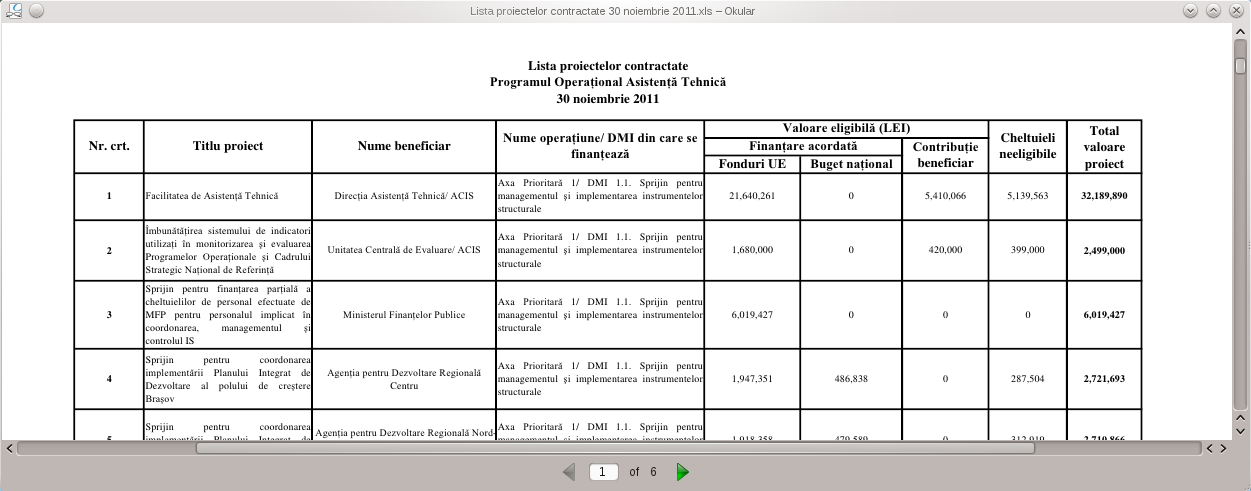}
  \caption{Example of publication of Romanian ERDF spending data (PDF
    file)}
  \label{fig:erdf_ro}
\end{figure*}

\section{Introduction and Background}

Open data is an approach to data management based on the tenet that
``certain data should be freely available to everyone to use and
republish'' (Wikipedia). Open data has increasingly gained traction
over the years and by now is supported by parts of academia,
government and business.

Open data can be characterized as information processing with the goal
to create knowledge and to manipulate that knowledge effectively
(e.g., via collaborative tagging and interactive mash-up
visualizations). Arguably, the idea of open data was first brought to
the attention of a broader audience with an article from UK's The
Guardian in 2006, which had the opening line: ``Our taxes fund the
collection of public data -- yet we pay again to access it.  Make the
data freely available to stimulate innovation'' \cite{AC:Guardian:06}.
However, it should be noted that efforts started earlier than
that---the Australian government has been moving towards more open
data management since at least 2001 \cite{GTF:Engage:09}.

Besides open data, there are related concepts such as open content
(\url{opencontent.org/definition/}), open access
(\url{www.earlham.edu/~peters/fos/overview.htm}) and open knowledge
(\url{opendefinition.org/okd/}), but their boundaries are blurry; in
the following we use only the term open data with the understanding
that it should be interpreted in a broad sense.

In academia there is the recognition that scientific data should be
freely available to speed up scientific advances and to enable new
forms of collaborative research (e.g., science 2.0 and open notebook
science \cite{BOB:FirstMonday:10}). In government, data is made
available to increase transparency how government operates and to
encourage participation of citizens. Open data in the governmental
domain is encouraged by laws such as the European Directive on the
Re-Use of Public Sector Information (PSI Directive) and the Freedom of
Information Act (FOIA) in the US. The Obama administration pursues
the Open Government Initiative to ``ensure the public trust and
establish a system of transparency, public participation, and
collaboration'' (\url{www.whitehouse.gov/open}) while the Digital
Agenda for Europe calls for action to ``open up public data sources
for re-use'' (\url{ec.europa.eu/information_society/digital-agenda}).

One can argue for open data from many angles, including societal and
economic benefits; conversely, there are also concerns such as
potential privacy risks and the fear that raw data can be
misinterpreted \cite{LAY:HBS:10} \cite{MEPSIR:06} \cite{Weiss:UnPbl:02}
\cite{DataGov:ConOps}. Regardless of its perceived potential and
risks, it is a fact that increasingly data is made available in an
open manner. This trend is also apparent by emerging events such as
the Open Government Data Camp (\url{ogdcamp.org/}) and the Open
Knowledge Conference (OKCon) (\url{okcon.org}).

Open data is already reality. The UK government has made available so
far more than 7,000 datasets at \url{data.gov.uk}. Other examples of
dataset providers are the Open Knowledge Foundation's
\url{publicdata.eu}, the US government (\url{www.data.gov}, over 3,700
datasets), and The World Bank (\url{data.worldbank.org/data-catalog},
over 7,000 datasets).  Furthermore, states, regions, and cities have
also started open data initiatives; to give one of many examples, the
city of Munich has started to publish information as open data and has
held the Munich Open Government Day
(\url{www.muenchen.de/Rathaus/dir/limux/mogdy/Programmierwettbewerb/}).

The Open Government Data (OGD) Stakeholder Survey
(\url{survey.lod2.eu}) conducted in 2010 has collected 329 responses
from citizens, politicians, public administrators, industry, media and
science that are producers, publishers and/or consumers of open data
\cite{MKNA:OKCon:11}.
The survey revealed that national datasets are most desirable before
regional and worldwide ones and that important (quality) criteria for
open data are provenance/source of data, format, completeness of
metadata, and official certificates. Users of open data are most
interested in geospacial, economic and financial data and want to do
research/analysis, visualization, and simply consuming of the data.

It is expected that open data will continue to be implemented by a
growing number of governments and organizations. Thus, the handling of
open data will increase and with it the need to have effective tools
and techniques to manage and maintain it.
In this paper we argue that software maintenance and reverse
engineering has an opportunity to contribute to open data and to shape
its future development. The baseline for this observation is that from
the viewpoint of reverse engineering open data is just another new
artifact as input to the reverse engineering process. Reverse
engineering has continuously broadened its artifacts going beyond
source code and databases \cite{MK:EoSE:10} to, for instance, images
(CAPCHAs) \cite{HGH:WCRE:08} and (business) processes \cite{TZ:WSE:10}
\cite{HGH:ICSM:10}. Of course, all these artifacts can be treated as
data (including source code). 

Similarly, open data and its infrastructure has several maintenance
challenges that need to be studied so that domain-specific techniques
and tools can be developed to meet key requirements such as
verifiability and traceability. The mining of software artifacts and
their interdependencies \cite{HMHJ:TSE:05} can be extended and adapted
towards open data with the goal to, for instance, improving on
detecting and correction of ``buggy'' data items and data
extractors/transformers, and studying of open data maintenance
processes and collaboration patterns among different groups of
contributors (e.g., people concerned with data scraping,
manipulating/abstracting and visualizing).




The reminder of the paper is organized as follows.
Section~\ref{sec:example} provides a real-world example (ERDF data) to
illustrate the current state of open data and its challenges. ERDF
data is distributed over various locations using different formats and
inconsistent meta-data. Other examples of open data exhibit similar
challenges.
Drawing from this example, Sections~\ref{sec:re} and
\ref{sec:maintenance} discuss the reverse engineering and maintenance
perspectives of open data, respectively. For each perspective we
identify challenges and research opportunities.
Section~\ref{sec:concl} concludes the paper.

\section{Illustrative Example}\label{sec:example}

The European Regional Development Fund (ERDF) distributes money to
regions in Europe with the objective ``to help reinforce economic and
social cohesion by redressing regional imbalances''
(\url{europa.eu/legislation_summaries/agriculture/general_framework/g24234_en.htm}).
Its current funding round runs from 2007--2013, has a budget of EUR 201
billion, and is governed by various regulations. The implementing
regulation, Commission Regulation (EC) No 1828/2006, states in Article
7 that ``the managing authority shall be responsible for \dots{} the
publication, electronically or otherwise, of the list of
beneficiaries, the names of the operations and the amount of public
funding allocated to the operations.'' This requirement has been newly
introduced in a push towards increasing transparency.
As a consequence, the managing governmental authorities of ERDF
funds typically make this information available on public Web sites.

The European Commission maintains a collection of links that point to
the individual data sources
(\url{ec.europa.eu/regional_policy/country/commu/beneficiaries/index_en.htm}).
Depending on the country, there can be a single, centralized access
point or multiple access points of a country's (groups of) regions,
provinces, states, etc. For example, Romania has a central site, each
German state maintains its own Web site, and The Netherlands has four
Web sites, each encompassing several provinces.

\begin{figure*}[tb]
  \centering
  \includegraphics[width=1.5\columnwidth]{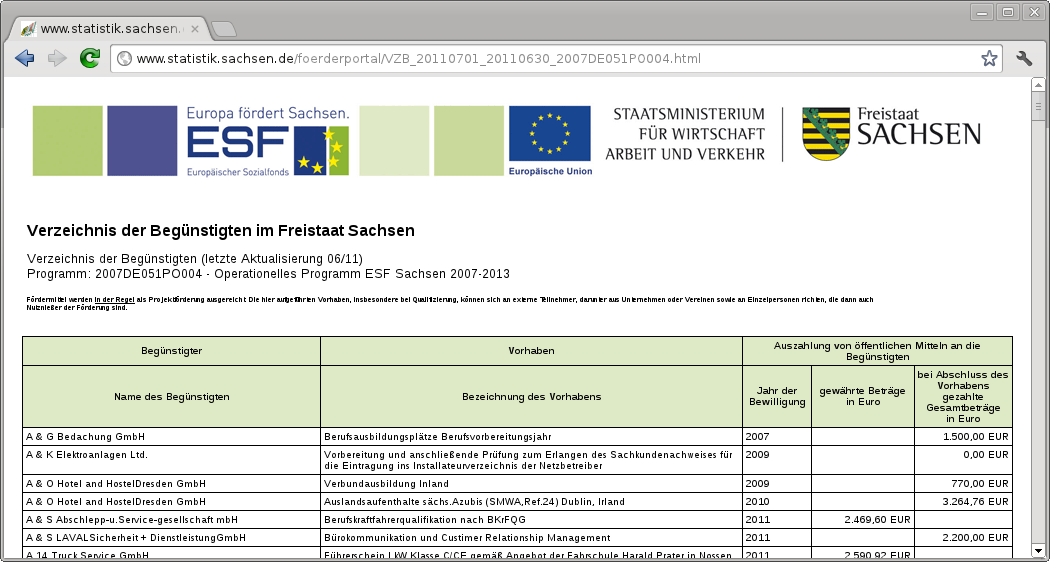}
  \caption{Publication of ERDF spending for Saxony (single HTML
    table)}
  \label{fig:erdf_sachsen}
\end{figure*}

\begin{figure}[tb]
  \centering
  \includegraphics[width=\columnwidth]{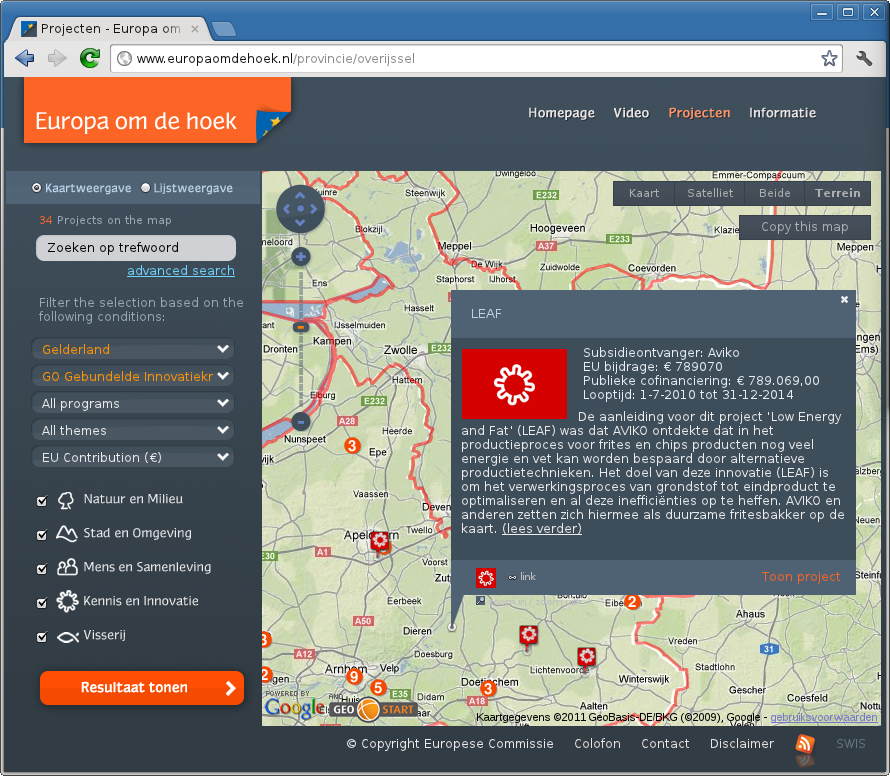}
  \caption{ERDF spending for The Netherlands (Web app)}
  \label{fig:leaf}
\end{figure}

The Romanian site (\url{www.fonduri-ue.ro/proiecte-contractate-236})
publishes data monthly in a RAR archive that contains a set of seven
PDF files. Figure~\ref{fig:erdf_ro} gives an example how the PDF is
organized.
The site of the German state of Saxony
(\url{www.statistik.sachsen.de/foerderportal/}) provides a single HTML
table of all spending data ordered by the name of the beneficiary (cf.
Figure~\ref{fig:erdf_sachsen}). Besides having four separate sites for
different groups of regions, The Netherlands has a dedicated site
(\url{www.europaomdehoek.nl}) that provides a Web application for
interactive exploration (cf.  Figure~\ref{fig:leaf}).

Open data activists have the goal to collect, abstract and visualize
all ERDF data in a consistent manner. The Financial Times and the
Bureau of Investigative Journalism did work on a consolidated spending
database in 2010 because ``there has been little transparency about
how the funds are used'' \cite{O'Murchu:OKFN:11}. They found a number
of misuses and abuses of funds that let them to conclude ``the
concepts of what EU representatives think of as transparency and what
actually allows citizens to easily understand how the 27-member bloc
spends the Structural Funds are worlds apart.''

\begin{figure}[tb]
  \centering
  \fbox{\begin{minipage}[h]{.95\columnwidth}
      \small \input{schema.sql}
    \end{minipage}
  }
  \caption{Schema of ERDF database}
  \label{fig:erdf_schema}
\end{figure}

A query interface to the database is available at
\url{eufunds.ftdata.co.uk/}. As a single zipped file in SQLite format
the database is about 600MB.\footnote{Personal email communication
  with Friedrich Lindenberg (\url{pudo.org}).} The database has a flat
schema (i.e., a single table; cf. Figure~\ref{fig:erdf_schema}) so
that it be can easily mapped to spreadsheet/CSV formats. All fields in
the schema are text and there are many entries that are not directly
available from the data sources.

Constructing the consolidated database was a major effort because
``the data were published on more than 100 websites, in nearly 600
documents and in 21 languages. So, while the information was, in
principle, freely available, it was not presented in a way that could
be meaningfully analyzed'' \cite{O'Murchu:OKFN:11}. Also, data was not
always available (Greece published blank tables in PDF files),
incomplete (Belgium), outdated (some German states), wrong (UK), or
password protected \cite{O'Murchu:OKFN:11} \cite{Barr:BoIJ:10}.


For the next funding round (2014--2020) the European Commission is
working on regulations to encourage open data such as a centralized
database that contains more project details (e.g., EU co-financing
rate and total spending) and adheres to the \emph{8 Principles of Open
  Government Data} (\url{www.opengovdata.org/home/8principles}).




\section{Reverse Engineering Perspective}\label{sec:re}

A prerequisite for open data is to obtain (raw) data in a form such
that it can be effectively used for information processing and
visualization, knowledge generation and decision making.
Unfortunately, even if data is accessible it still needs to be
transformed to enable its (effective) use. This step is essentially
very similar to traditional software reverse engineering---using
Chikofsky and Cross's classical definition \cite{CC:IEESW:90} as a
baseline, we can define reverse engineering for open data as the
creation of data representations that (1) transform the original data
to another form and/or (2) transform it into a higher level of
abstraction. Note that the original data is not changed; in fact, it
often resides at an authoritative source that provides read-only
access.

Transforming of the original data into another form is typically
required because the data is not efficiently machine-readable,
queryable or storable. Open data is made available in many different
formats such as XML, HTML, Word documents, Excel spreadsheets,
comma-separated values format (CSV), and PDF files.
According to the OGD Stakeholder Survey the most popular current
formats are HMTL (52\%), PDF (50\%), CSV/XLS (37\%), DOC/RTF (32\%),
XML (27\%), APIs (22\%) and RDF (18\%) \cite{MKNA:OKCon:11}. Thus,
formats that are easily machine processable are currently loosing out
to other formats.  Indeed, according to the survey the most requested
(future) formats are APIs, XML and RDF.


\begin{figure*}[tb]
  \centering
  \includegraphics[width=.99\textwidth]{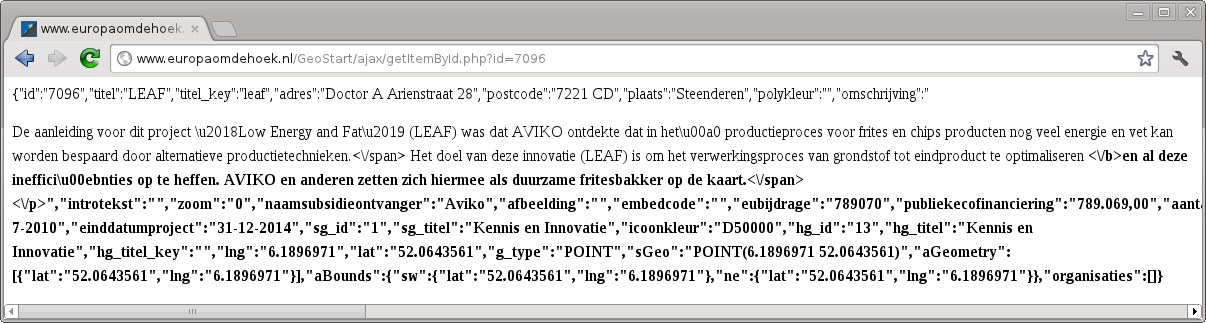}
  \caption{Example of raw API access for ERDF data of The Netherlands}
  \label{fig:leaf_json}
\end{figure*}


For open data publishing the following quality levels can be
distinguished (based on Shadbolt \cite{Shadbolt:ISWC:10}):
\begin{description}[\IEEEsetlabelwidth{addressable:}]
\item[available:] data is accessible on the Web (in any form or
  format)
\item[structured:] data is structured (e.g., CSV and Microsoft Office
  binary formats)
\item[standardized:] data uses open, standardized formats (e.g., XML,
  RDF and JSON)
\item[addressable:] individual data-points are denoted by a unique URL
\item[linked:] data links to data from external sources (other data
  providers)
\end{description}

In order to process data at the lower levels (available, structured,
and standardized) some kind of reverse engineering is typically
required. At this point in time, open data is accessible mostly at the
two lowest levels (available and structured). ERDF data (cf.
Section~\ref{sec:example}) is almost exclusively published as PDF or
HTML, and ``no data is currently published in XML or JSON or RDF''
\cite{Reggi:blog:11}.

Even if data is structured, the encoding may vary (e.g., for CSV
different conventions are used to denote field/record separators,
string entities, date and time, and so on).  If data is structured,
the data's schema (or metadata) may not be available. In this case, it
may be desirable to (semi-)automatically recover the data's structure
(\emph{schema recovery}).


In practice, processing of PDF files are a major concern. They are
very inconvenient to process while being surprisingly common in
practice.  For example, all departments of the UK government publish
their annual reports as PDFs and ``those PDFs are full of tables,
however not one department publishes these as a spreadsheet or any
accessible format'' \cite{Rogers:Guardian:11}. PDF is a complex
format\footnote{The ISO standard 32000-1:2008 that covers version 1.7
  of the PDF format has almost 800 pages
  (\url{www.adobe.com/devnet/acrobat/pdfs/PDF32000_2008.pdf}).}  that
can include PostScript and JavaScript code, forms, and vector/raster
images; so far there are nine different official versions of PDF with
differing capabilites.
Also, a PDF may contain features that are only supported by Adobe
software and that cannot be processed by other PDF viewers. As a
result each PDF file may pose different challenges when extracting its
content. This causes many practical problems, for instance in a PDF
containing the spending of a UK department ``the core tables were
impossible to export'' and for another department ``the tables were so
badly formatted in the original PDFs that we had to copy the data out
by hand'' \cite{Rogers:Guardian:11}.

In the following subsections we briefly outline reverse engineering
challenges and techniques that are needed for open data.

\subsection{Scraping}

Open data is typically made available on the Web.  Often there is a
permanent URL that points to a self-enclosed document, but it is also
common that data is embedded within static HTML or a dynamic Web
application. ScraperWiki (\url{www.scraperwiki.com}) is an example of
a portal that allows to develop and run scripts in Python, Ruby and
PHP. Scripts scrape Web sites that contain open data and make the
results available for simple interactive exploration or for download
(CSV, JSON, or SQLite). ScraperWiki scripts can use libraries that
simplify processing (e.g., lxml.html for HTML parsing).

An example of a static HTML page is the ERDF data for Saxony. There is
a ScraperWiki
script\footnote{\url{scraperwiki.com/scrapers/eu_regional_development_fund_recipients_-_saxony_g/}}
(37 lines of Python code) that processes this data. As typical for
such a scrapers, the script would break if layout and/or names change
in the HTML encoding.
Another similar example is the WHO's Global Alert and Response
information that can be obtained for the years 1996--2011 with
differently URLs:
\url{www.who.int/csr/don/archive/year/}\textit{yyyy}\url{/en/}. The
ScraperWiki script (\url{scraperwiki.com/scrapers/who_outbreaks/})
that processes this data is 63 lines of Python code.


The ERDF Web application for The Netherlands has neither static HMTL
nor an official API to obtain the data. Via reverse engineering of the
Web application (e.g., with the help of a JavaScript debugger) a
query-URL can be obtained, which takes a project ID and in return
provides the raw data for the corresponding project in JSON.
Figure~\ref{fig:leaf_json} shows an example of the query-URL with
project ID 7096 (which provides the raw data for the visualization in
Figure~\ref{fig:leaf}).\footnote{A ScraperWiki script can be found at
  \url{scraperwiki.com/scrapers/dutch_european-funded_regional_development_project/}.}
For Web sites that provide a query interface only, it can be difficult
or impossible to determine the size of the underlying database and to
assure exhaustive extraction of the available data. In such cases a
semi-automatic approach for filling out search queries is desirable.
Interestingly, this problem is also encountered by search engines that
have to cope with the so-called \emph{hidden Web} \cite{CHM:CACM:11}.



\subsection{Image Processing}

Reverse engineering of open data can require the transformation of
bitmaps towards characters and vector data. This typically entails
optical character recognition (OCR). But layout and lines may also
need to be processed, for instance for tables or multi-column text,
which can be handled by \emph{document image analysis} \cite{OK:09}.
An example that requires this approach is the ERDF data of Bulgaria,
which is provided as bitmaps embedded in PDFs.

In the reverse engineering literature there are examples of image
processing techniques and OCR in the areas of CAPCHAs
\cite{HGH:WCRE:08}, UML diagrams \cite{LTJ:CASCON:00} and GUI testing
\cite{CVO:ICSM:10} that might be applicable for the processing of open
data as well.
For instance, GUI testing of a Web site for different browsers can be
accomplished by ``graphical diffing'' of the rendered pages.
Similarly, table structures of different PDF documents could be
graphically differenced.

There are generic tools available that provide functionality for
converting PDFs to text such as Adobe Reader and \texttt{pdftotext}
(part of Xpdf). However, depending on the complexity and
PDF-representation of information its structure can get lost. For
example, when Acrobat Reader 9 extracts content for the PDF in
Figure~\ref{fig:erdf_ro} text is in the wrong order and column
boundaries are lost. The \texttt{pdftotext} tool provides a much more
usable extraction for this PDF file, but the table header is not
correctly recognized.

A general problem is that converters are not customizable. It would be
desirable, for instance, to specific table layouts also with the help
of graphical regions. Imagine the typical scenario of a PDF file that
contains a single table spread over many pages. If the columns of the
table are consistently located at the same horizontal offsets a
geometric specification could be easily used as guidance for data
extraction.

\subsection{Structure and Schema Recognition}

Open data may be made available as a single table without much
structure. As the ERDF database (cf. Figure~\ref{fig:erdf_schema}))
illustrates, there can be a large number of fields/columns with many
rows of data.

From a database perspective, this data is only in first normal form
(1NF). While this format permits SQL-style queries, it has little
structure. Field information needs to be repeated on each row. For
example, in the ERDF database if a certain beneficiary has multiple
projects then all of the beneficiary's information is repeated,
possibly with variations (e.g., with or without diacritical marks or
different capitalization). Such variations can introduce mistakes when
data is transformed and consolidated.

Open data is typically published without a description of the schema,
formally or informally. The meaning of ``schema'' should be
interpreted broadly in the sense that Word-style and HTML documents
can have structures as well (e.g., a certain combination of font
attributes could have a certain meaning).
There are \emph{hypertext-based data models} ``in which page authors
use combinations of HTML elements (such as a list of hyperlinks),
perform certain data-model tasks (such as indicate that all entities
pointed to by the hyperlinks belong to the same set)''
\cite{CHM:CACM:11}.

For such data as well as for collaboratively constructed and mashed-up
open data one cannot assume ``centralized data design'' as it is found
at traditional databases. Reverse engineering techniques could be used
to recover and complete schema information and to infer
constraints/structures of the data.  There is promising research in
that direction for Web-embedded structured data \cite{CHM:CACM:11}
\cite{VHMPS:VLDB:11}. Another example is the OpenII open source tool
set (\url{openii.sourceforge.net/}) for data integration tasks such as
clustering and visualization of schemas as well as matching of
source/target schemas.


\begin{figure*}[bt]
  \centering
  \includegraphics[width=1.5\columnwidth]{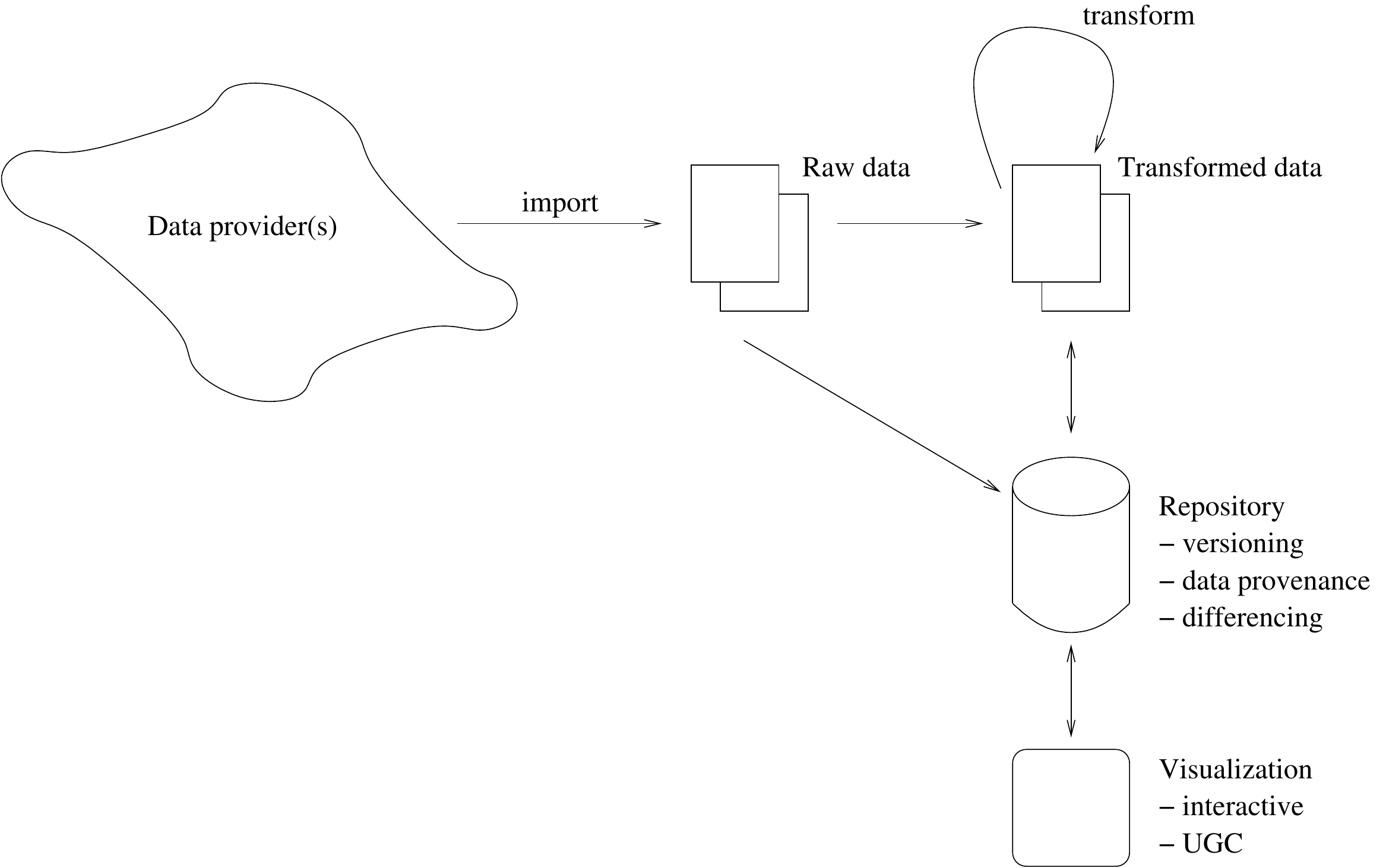}
  \caption{Model of open data processing}
  \label{fig:bigpic}
\end{figure*}

\section{Maintenance Perspective}\label{sec:maintenance}

While the reverse engineering perspective is mostly concerned with the
transformation of information in individual documents/databases, the
maintenance perspective addresses the management and flow of
information as well as quality attributes of the whole process.

We propose the following model of open data processing.
Figure~\ref{fig:bigpic} gives an overview of the model with the most
important elements and the data flow among them. The output of
processing is to present information in a novel form that allows to
gain unique insights that would not have been possible with the raw
data sources. Typical examples are (interactive) visualizations that
provide abstractions and expose dependencies of data.
The inputs to the process are multiple, diverse data sources. These
data sources are typically from independent third parties. Thus, data
needs to be pulled from sources and it is expected that the data
provider may modify data in the future in a manner that is more or
less unpredictable for the consumer. Data is kept in a (central)
repository, which needs to support versioning, data provenance and
differencing (discussed below). This model shares similarities with
the ones that are found in reverse engineering (extract, abstract and
present) and data warehousing (extract, transformation and load
(ETL)).

The data processing is organized as a transformation
pipeline.\footnote{Instead of a linear pipeline one can image a more
  complex model based on \emph{flow graphs} where nodes in the graph
  represent operators, which can be flow manipulations or
  transformations. The SPADE programming language for stream
  processing is based on this principle \cite{DePauw:RV:10}.} A
transformation step may, for instance, (1) change the format, data
representation, and/or schema, (2) augment data (e.g., aggregation of
data items or annotation/cross-referencing of data items), and (3)
perform sanity checks and validate (schema) constraints.

Note that each transformation may be manual, semi-automatic or
fully-automatic. Since data sources are often only semi-structured,
human verification and corrections are not uncommon. For instance, the
OCR recognition of a number may we wrong for a certain data item. Once
this is recognized (via manual inspection or violation of a sanity
constraint) a hand-written transformation could be added to the
transformation pipeline to fix this data item.

To analyze the (transformed) data in an effective manner by its users
visualizations are needed. Visualizations can be text-based,
graphical, or both and are typically made available as a web
interface. Visualizations need a query interface to the repository
(which may differ from the way that transformations access the
database). Visualizations may support user-generated content (UGC)
that enhances the ``baseline'' repository with additional knowledge
(e.g., URIs to external data sources).

Since the outputs of the process are expected to be used by research,
businesses and governments to advance their understanding, trust in
the data and transparency in the processing is essential. In this
context, key requirements to support are versioning and traceability
for quality control. A closely related research field is \emph{data
  provenance}, which can be defined as ``information that helps
determine the derivation history of a data product, starting from its
original sources;'' and furthermore ``the two important features of
the provenance of a data product are the ancestral data product(s)
from which this data product evolved, and the process of
transformation of these ancestral data product(s), possibly through
workflows, that helped derive this data product'' \cite{SPG:IUB:05}.

For each data item at the output it should be possible to trace its
dependencies through the transformations back to the source data. This
is important for debugging and assurance.  All data sources,
transformations, etc. need to be versioned so that output can be
faithfully reproduced later on if needed. If a data provider makes a
modification (e.g., change of an existing PDF file) or addition (e.g.,
a new PDF becomes available) it needs to be properly versioned.

If a data source has been modified there needs to be effective tools
support to analyze the differences (``deltas''). It may be that the
underlying information has changed, that the representations or
encoding has changed, or both. Think of a PDF file that looks
identical to the eye, but whose encoding has changed such that
\texttt{pdftotext} produces now different output that breaks
assumptions in the transformation pipeline.
Generally, it is desirable to be able to analyze deltas for two (or
more) configurations of runs.

Since open data infrastructure is just starting to emerge there are no
dominant technologies and infrastructures yet. Once they are emerging
one can expect that projects will have to be migrated to more
established platforms (i.e., both data and software migration). The
W3C's SPARQL (\url{www.w3.org/TR/rdf-sparql-query/}) is currently
discussed as a possible query end point for open data
\cite{EM:OpenLink:09}.\footnote{A collection of SPARQL endpoints for
  open data is available at
  \url{labs.mondeca.com/sparqlEndpointsStatus/}.} To accomplish this,
Web applications such as the ERDF app from The Netherlands (cf.
Section~\ref{sec:example}) will have to be migrated towards a SPARQL
API.

\section{Conclusions}\label{sec:concl}

In this paper we have outlined the push for open data and described
its current state with the help of an example---open data for the
beneficiaries of European Regional Development Fund money. We then
described research challenges for open data in the areas of reverse
engineering and maintenance. Open data presents not only worthwhile
research opportunities, but promises to benefit society as
well.
It is our hope that this paper will inspire other researchers within
the reverse engineering and maintenance communities to take up the
open data challenge.


\bibliographystyle{IEEEtran}
\bibliography{papers,own,books,local}

\begin{thebibliography}{10}

\bibitem{AC:Guardian:06}
Charles Arthur and Michael Cross.
\newblock Give us back our crown jewels.
\newblock {\em The Guardian}, March 2006.
\newblock
  \url{http://www.guardian.co.uk/technology/2006/mar/09/education.epublic/print}.

\bibitem{Barr:BoIJ:10}
Caelainn Barr.
\newblock Lack of transparency hides details of funding.
\newblock {\em The Bureau of Investigative Journalism}, November 2010.
\newblock \url{http://www.thebureauinvestigates.com/2010/11/29/top-story-1/}.

\bibitem{BOB:FirstMonday:10}
Jean–Claude Burgelman, David Osimo, and Marc Bogdanowicz.
\newblock Science 2.0 (change will happen \dots).
\newblock {\em First Monday}, 15(7), July 2010.
\newblock
  \url{http://www.uic.edu/htbin/cgiwrap/bin/ojs/index.php/fm/article/view/2961/2573}.

\bibitem{CHM:CACM:11}
Michael~J. Cafarella, Alon Halevy, and Jayant Madhavan.
\newblock Structured data on the {Web}.
\newblock {\em Communications of the ACM}, 54(2):72--79, February 2011.

\bibitem{CC:IEESW:90}
Elliot~J. Chikofsky and James~H. Cross, II.
\newblock Reverse engineering and design recovery: A taxonomy.
\newblock {\em IEEE Software}, 7(1):13--17, January 1990.

\bibitem{CVO:ICSM:10}
Shauvik~Roy Choudhary, Husayn Versee, and Alessandro Orso.
\newblock {WebDiff}: Automated identification of cross-browser issues in web
  applications.
\newblock {\em 26th IEEE International Conference on Software Maintenance (ICSM
  2010)}, October 2010.

\bibitem{DePauw:RV:10}
Wim De~Pauw.
\newblock {\em 1st International Conference on Runtime Verification (RV 2010)},
  volume 6418 of {\em Lecture Notes in Computer Science}, chapter Visual
  Debugging for Stream Processing Applications, pages 18--35.
\newblock Springer-Verlag, 2010.

\bibitem{MEPSIR:06}
Makx Dekkers, Femke Polman, Robbin {te Velde}, and Marc {de Vries}.
\newblock {MEPSIR}: Measuring european public sector information resources.
\newblock
  \url{http://ec.europa.eu/information_society/policy/psi/actions_eu/policy_actions/mepsir/index_en.htm},
  June 2006.

\bibitem{EM:OpenLink:09}
Orri Erling and Ivan Mikhailov.
\newblock {SPARQL} and scalable inference on demand, 2009.
\newblock
  \url{http://www.openlinksw.com/weblog/oerling/scalable_inference.pdf}.

\bibitem{GTF:Engage:09}
{Government 2.0 Taskforce}.
\newblock Engage: Getting on with government 2.0.
\newblock \url{http://www.finance.gov.au/publications/gov20taskforcereport/},
  2009.

\bibitem{HMHJ:TSE:05}
Ahmed~E. Hassan, Audris Mockus, Richard~C. Holt, and Philip~M. Johnson.
\newblock Guest editor's introduction: Special issue on mining software
  repositories.
\newblock {\em IEEE Transactions on Software Engineering}, 31(6):426--428, June
  2005.

\bibitem{HGH:WCRE:08}
Abram Hindle, Michael~W. Godfrey, and Richard~C. Holt.
\newblock Reverse engineering {CAPTCHAs}.
\newblock {\em 15th IEEE Working Conference on Reverse Engineering (WCRE'08)},
  pages 59--68, October 2008.

\bibitem{HGH:ICSM:10}
Abram Hindle, Michael~W. Godfrey, and Richard~C. Holt.
\newblock Software process recovery using recovered unified process views.
\newblock {\em 26th IEEE International Conference on Software Maintenance (ICSM
  2010)}, October 2010.

\bibitem{LAY:HBS:10}
Karim~R. Lakhani, Robert~D. Austin, and Yumi Yi.
\newblock Data.gov.
\newblock Case Study 9-610-075, Harvard Business School, May 2010.
\newblock \url{http://www.data.gov/documents/hbs_datagov_case_study.pdf}.

\bibitem{LTJ:CASCON:00}
Edward Lank, Jeb~S. Thorley, and Sean {Jy-Shyang Chen}.
\newblock An interactive system for recognizing hand drawn uml diagrams.
\newblock {\em Conference of the Centre for Advanced Studies on Collaborative
  Research (CASCON'00)}, October 2000.

\bibitem{MKNA:OKCon:11}
Michael Martin, Martin Kaltenb{\"o}ck, Helmut Nagy, and S{\"o}ren Auer.
\newblock The open government data stakeholder survey.
\newblock {\em 6th Open Knowledge Conference (OKCon'11)}, June 2011.
\newblock \url{http://ceur-ws.org/Vol-739/paper_6.pdf}.

\bibitem{MK:EoSE:10}
Hausi~A. M{\"u}ller and Holger~M. Kienle.
\newblock {\em Encyclopedia of Software Engineering}, chapter Reverse
  Engineering, pages 1016--1030.
\newblock Taylor \& Francis, 2010.
\newblock \url{http://www.informaworld.com/10.1081/E-ESE-120044308}.

\bibitem{DataGov:ConOps}
{Office of E-Government and IT Office of Management and Budget}.
\newblock {Data.gov} concept of operations.
\newblock Version 1.0,
  \url{http://www.data.gov/documents/data_gov_conops_v1.0.pdf}.

\bibitem{OK:09}
Lawrence O'Gorman and Rangachar Kasturi.
\newblock {\em Document Image Analysis}.
\newblock Reissued after out of print, 2009.
\newblock
  \url{citeseerx.ist.psu.edu/viewdoc/download?doi=10.1.1.182.6107&rep=rep1&type=pdf}.

\bibitem{O'Murchu:OKFN:11}
Cynthia O'Murchu.
\newblock A {Kafkaesque} data-trail: the hunt for `{Europe's} hidden billions'.
\newblock {\em Open Knowledge Foundation Blog}, March 2011.
\newblock
  \url{http://blog.okfn.org/2011/03/08/a-kafkaesque-data-trail-the-hunt-for-europes-hidden-billions/}.

\bibitem{Reggi:blog:11}
Luigi Reggi.
\newblock The map of {EU} structural funds transparency at regional level -
  october 2011.
\newblock {\em Regional Innovation Policies}, October 2011.
\newblock
  \url{http://www.luigireggi.eu/Innovation-policies/Home/Entries/2011/10/31_The_map_of_EU_Structural_Funds_Transparency_at_regional_level_-_October_2011.html}.

\bibitem{Rogers:Guardian:11}
Simon Rogers.
\newblock Named and shamed: the worst government annual reports.
\newblock {\em The Guardian: Data Blog}, October 2011.
\newblock
  \url{http://www.guardian.co.uk/news/datablog/2011/oct/27/department-resource-accounts-reports}.

\bibitem{Shadbolt:ISWC:10}
Nigel Shadbolt.
\newblock Why open government data: Lessons from data.gov.uk.
\newblock {\em International Semantic Web Conference (ISWC'10)}, November 2010.
\newblock \url{http://eprints.ecs.soton.ac.uk/21648/}.

\bibitem{SPG:IUB:05}
Yogesh~L. Simmhan, Beth Plale, and Dennis Gannon.
\newblock A survey of data provenance techniques.
\newblock Technical Report IUB-CS-TR618, Computer Science Department, Indiana
  University, Bloomington, August 2005.
\newblock \url{http://www.cs.indiana.edu/pub/techreports/TR618.pdf}.

\bibitem{TZ:WSE:10}
Ran Tang and Ying Zou.
\newblock An approach for mining web service composition patterns from
  execution logs.
\newblock {\em 12th IEEE International Symposium on Web Systems Evolution (WSE
  2010)}, pages 53--62, September 2010.

\bibitem{VHMPS:VLDB:11}
Petros Venetis, Alon Halevy, Jayant Madhavan, Marius Pasca, Warren Shen, Fei
  Wu, Gengxin Miao, and Chung Wu.
\newblock Recovering semantics of tables on the web.
\newblock {\em 37th International Conference on Very Large Data Bases
  (VLDB'11)}, pages 528--538, August 2011.
\newblock \url{http://ilpubs.stanford.edu:8090/1012/1/tables.pdf}.

\bibitem{Weiss:UnPbl:02}
Peter Weiss.
\newblock Borders in cyberspace: Conflicting public sector information policies
  and their economic impacts.
\newblock \url{http://www.weather.gov/sp/Borders_report.pdf}, February 2002.

\end{thebibliography}
\end{document}